# Pair-Variational Autoencoders (PairVAE) for Linking and Cross-Reconstruction of Characterization Data from Complementary Structural Characterization Techniques

Shizhao Lu[1], Arthi Jayaraman[1,2]*

**Affiliations**

[1] Department of Chemical and Biomolecular Engineering, University of Delaware, Newark, DE 19716

[2] Department of Materials Science and Engineering, University of Delaware, Newark, DE 19716

**ABSTRACT**

In materials research, structural characterization often requires multiple complementary techniques to obtain a holistic morphological view of the synthesized material. Depending on the availability and accessibility of the different characterization techniques (e.g., scattering, microscopy, spectroscopy), each research facility or academic research lab may have access to high-throughput capability in one technique but face limitations (sample preparation, resolution, access time) with other technique(s). Furthermore, one type of structural characterization data may be easier to interpret than another (e.g., microscopy images are easier to interpret than small angle scattering profiles). Thus, it is useful to have machine learning models that can be trained on paired structural characterization data from multiple techniques (easy and difficult to interpret, fast and slow in data collection or sample preparation), so that the model can generate one set of characterization data from the other. In this paper we demonstrate one such machine learning workflow, PairVAE, that works with data from Small Angle X-Ray Scattering (SAXS) that presents information about bulk morphology and images from Scanning Electron Microscopy (SEM) that presents two-dimensional local structural information of the sample. Using paired SAXS and SEM data of newly observed block copolymer assembled morphologies [open access data from Doerk G.S., et al. Science Advances. 2023 Jan 13;9(2): eadd3687], we train our PairVAE. After successful training, we demonstrate that the PairVAE can generate SEM images of the block copolymer morphology when it takes as input that sample's corresponding SAXS 2D pattern, and vice versa. This method can be extended to other soft materials morphologies as well and serves as a valuable tool for easy interpretation of 2D SAXS patterns as well as an engine for generating ensembles of similar microscopy images to create a database for other downstream calculations of structure-property relationships.

# INTRODUCTION

Machine learning and artificial intelligence are becoming invaluable methods that enable faster discovery, innovation, and automation in chemical and material sciences and engineering.[1-10] In particular, machine learning has found tremendous use in automating materials' structural analysis which is a vital step in establishing the design-structure-property relationship of a novel material.[11-21] For example, structural characterization of materials often depend of microscopy imaging techniques (e.g., scanning electron microscopy or SEM, transmission electron microscopy or TEM, atomic force microscopy or AFM) to visualize nanoscale or microscale structural features or patterns.[22] Deep learning models used for pattern recognition and image analysis have been adopted as a viable means to automatically extract structural from microscopy images information regarding the ordered arrangements of the molecules, type of ordered assembly, and detection of objects' (e.g., nanoparticles, assembled domains) shapes and sizes.[23-35] There are unique challenges, however, with training machine learning models that are used for everyday photographic image analysis to analyze materials' microscopy images. For example, compared to photographic images of everyday objects that often are easily recognizable, materials' microscopy images require detailed metadata of the material, chemistry, synthesis conditions, imaging process conditions as labels. The richness of metadata associated with each microscopy image also implies limitation on the curation of large microscopy databases because a generic (universal) labeling scheme would be nearly impossible for microscopy images coming from diverse sources. Furthermore, a unique problem to (non-biological) materials field is the smaller dataset sizes for the microscopy images collected from samples due to the laborious sample preparation required to acquire each image; this problem restricts the use of deep learning models that require large training datasets like those available in biomedical fields.[36, 37] To address this challenge with small materials' microscopy image datasets, Lu et al.[38] recently proposed a semi-supervised transfer learning workflow that successfully classified the materials' morphologies from transmission electron microscopy images of protein/peptide nanowires despite being trained with less than ten labeled images per morphology.

Another machine learning approach that can be applied to the small and highly chemistry/material/processing specific microscopy image datasets is *generative* machine learning. Generative machine learning models are trained to reconstruct outputs that resemble the inputs, where the input could be two-dimensional images, three-dimensional structure, graph of a chemical structure of a molecule, a x-y plot, etc. One specific type of generative machine learning model is an autoencoder that can learn reconstruction of complex input data and/or obtain concise, low-dimensional latent space representations of the complex input data. An autoencoder consists of an encoder that encodes information from the input data to a set of lower dimensional latent variables and a decoder that decodes information from a set of latent variables back to the same number of dimensions as the input data. An autoencoder is typically trained by minimizing the mismatch (i.e., loss) between the reconstructed data and the input data. Material scientists have applied autoencoders for multiple purposes such as reconstruction of experimental characterization data[39-46], molecular structure[47-50] or microscopy image[41, 46, 51, 52]; clustering[53, 54] and/or classification[42, 43, 45, 48] of the latent space representations; obtaining material design parameters[39, 40] or deriving order parameters[55-59] from latent space representations; molecular[47] or material[52] property optimization based on the latent space representations. Some of the studies mentioned here[43-48, 50, 52, 53, 57] used a modified version of the autoencoder called *variational autoencoder*[60] (VAE) which maps the encoded latent space to a multidimensional standard Gaussian distribution, that has the benefit of continuous latent space compared to the sparse latent space that one would get from the encodings of an unmodified autoencoder. In a recent study Yaman et al.[46] demonstrated a dual VAE machine learning workflow that can be used to couple the surface photonic response spectroscopy (SPR) profiles and SEM images of gold nanoparticles. They have shown that by correlating the latent space of one VAE trained on SPR with the latent space of another VAE trained on SEM, they can correctly generate (or construct) SPR profile from a corresponding SEM image or vice versa.

In this article, partly inspired by the work of Yaman et al., we present a machine learning workflow called 'PairVAE' that can enable cross-reconstruction and cross-generation of complementary characterization data of soft materials. As a proof-of-concept, we demonstrate that PairVAE can reconstruct either Small Angle X-ray Scattering (SAXS) patterns or SEM images of block copolymer morphologies from the other type of characterization data with a limited number of training data of pairs of SAXS and SEM. For interested readers, in **Table. S1** we compare and contrast our PairVAE work (in terms of purpose, data domain, dataset format, model training protocol) against the work of Yaman et al.[46] that served as an inspiration to us. As our work in this paper uses block copolymer morphologies as an example for demonstrating PairVAE, we discuss briefly about block copolymers and the types of information one obtains by conducting SAXS and SEM measurements on them.

Block copolymers (BCPs) are polymers made up of two or more types of monomers with repeating sequences of different monomers. The self-assembly or directed assembly of BCPs in films or in solutions generates rich ordered morphologies including spheres, cylinders, lamellae, gyroids, vesicles, and other bicontinuous structures.[61-64] For melts of a simple linear diblock copolymer with two immiscible monomer chemistries, past studies using theory, simulations, and experiments have all shown self-assembly into four canonical - lamella, cylinder, spheres, and gyroid – ordered BCP morphologies by varying BCP composition and/or temperature or Flory-Huggins $\chi$ parameter (an indicator of immiscibility of the two monomers in the BCP).[65-67] In experiments, these BCP morphologies are most commonly characterized by complementary experimental techniques: Small-Angle X-ray or Neutron Scattering (SAXS or SANS) and Transmission Electron Microscopy (TEM) or Scanning Electron Microscopy (SEM). The full two-dimensional SAXS pattern quantifies the degree of positional and orientational order present in the BCP morphology over a range of length scales (inverse wavevector $q$). The azimuthally averaged one-dimensional SAXS or SANS scattering intensity $I(q)$ versus scattering wavevector $q$ is used to quantify the BCP domain spacing. In contrast, SEM or TEM images provide two-dimensional visualization of BCP morphologies at specific length scales. Unlike the SAXS and SANS patterns (2D or 1D) that provide average structural information of the bulk sample, SEM or TEM images for the same material can differ from sample to sample due to selection of the imaging region, differences of magnification/contrast, and existence of defects in the domains over the sample. While SAXS has the benefit of outputting statistical average or bulk structure information over multiple length scales, the data in wavevector $q$ space is abstract and requires non-trivial work in interpretation using analytical models[68, 69] or advanced computational techniques[15, 19, 70]. Unlike SAXS or SANS profiles that are harder to analyze and prone to incorrect interpretation if one fits with analytical models that are either approximate or inapplicable for the structure at hand, the SEM or TEM images are real-space visualization and less prone to incorrect interpretation. Thus, having both sets of results can provide unambiguous interpretation of the structure in the BCP system at various length scales. One could in principle apply fast Fourier transform on an SEM image to obtain a SAXS-like scattering pattern albeit with worse statistics than that obtained from SAXS measurement; furthermore, generation of local real-space SEM morphology representations from a SAXS pattern has not been reported in literature. Therefore, having a machine learning model that can couple/pair complementary SAXS and SEM results will be valuable as it can overcome the individual limitations/challenges for either approach. We describe our machine learning workflow, 'PairVAE', for linking and cross-reconstruction of 2D SAXS pattern and SEM images that provide complementary information about soft materials' morphologies.

## METHOD

Our machine learning workflow 'PairVAE' for cross-reconstruction of two types of complementary characterizations, 2D SAXS pattern and SEM image, was trained on paired SAXS and SEM data of BCP morphologies reported recently by Doerk et al.[71] They used blends of two BCPs both with polystyrene (PS) and poly(methyl methacrylate) (PMMA) blocks but with varying composition; one BCP was a lamellae forming symmetric PS-*b*-PMMA and the other BCP was cylinder forming PMMA-minority PS-*b*-PMMA.[72] These BCP blends were assembled into unique morphologies using template-directed assembly by varying the width of the substrate pattern.[71] For each of these morphologies, the authors presented complementary characterization data in the form of 2D SAXS patterns and SEM images. **Fig. 1** shows representative 2D SAXS pattern and SEM image for each of the four morphologies they observed. "Hexagonal" morphology consists of cylinder-like domains (bright dots and dim vertical lines) and lamella-like domains (bright vertical lines) in the SEM images with regular domain spacing. The authors noted that the "skew", "alternating lines", and "ladders" morphologies were not seen in previous studies. "Skew" morphology consists of bilayers of lamella-like domains at the air-interface with cylinder-like domains at the template-interface. "Alternating lines" morphology consists of vertical intercalating lines of bright lamella-like domains and dim cylinder-like domains while "ladders" morphology consists of oblong shapes with lamella-like domains parallel to the template grating direction and cylinder-like domain orthogonal to the template grating direction. In their work, SAXS was the high-throughput technique while the SEM characterization was slower and laborious and considered a bottleneck in their production of this paired characterization data. Their open-access dataset is tremendously useful for training and testing of a machine learning workflow like our PairVAE that can cross-reconstruct SEM (bottleneck method but easier to interpret) images from SAXS (high-throughput method but harder to interpret) data, and vice versa.

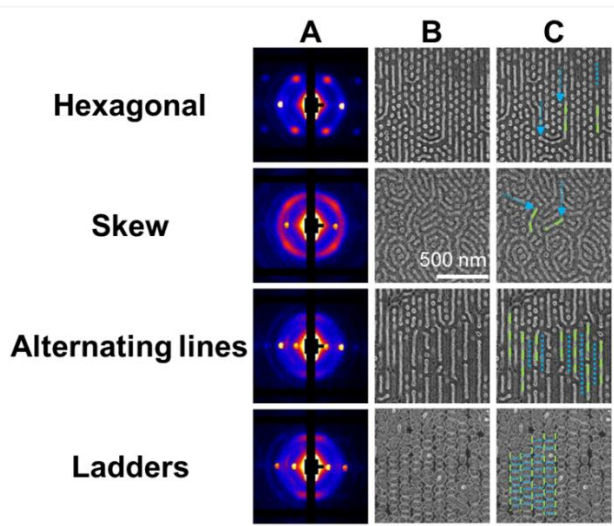

*Fig. 1. Representative characterization data of four different morphologies observed with a mixture of lamella-forming and cylinder-forming PS-PMMA BCPs assembled on a chemical grating template. The images in column **A**, column **B**, and column **C** are example SAXS pattern, randomly cropped SEM image, and annotated SEM image for each of the four morphologies. In the annotated images, green dashed lines indicate examples of lamella-like domains, blue arrows and dotted lines indicate cylinder-like domains. The white scale bar denoting 500 nm in ("skew" figure B) applies to all SEMs without scale bars. The full SEM images of the four morphologies are presented in **figure S1**. All of these SAXS and full SEM images were obtained from the open-access block copolymers characterization dataset by Doerk et al. Science*

'PairVAE' is comprised of two sets of variational autoencoders (VAE). Before the two sets of encoders and decoders in the two VAE are trained together to handle cross-reconstruction, each VAE is first trained on one type of characterization data through solo training as shown in **Fig. 2A**. We first describe briefly the dataset and the preprocessing of the characterization data used in these training steps.

We obtained 72 pairs of SAXS and SEM characterization data from their open-access data depository that were imaged at the same position in the larger grating template with the knowledge that the perception field of SAXS is larger than the SEM image. For SAXS data, we use the 2D SAXS pattern images resized to 192 × 192 pixels for training and reconstruction. For SEM data, the full SEM images are 830 × 1280 pixels with examples shown in **Fig. S1**, however, we do not use the full SEM images as our input for two reasons. One reason is that the different morphologies are visually distinguishable at smaller regions than the full SEM but less distinguishable with the full SEM image. The second reason is that the convolutional neural network (CNN) models that we use to process images work well with image size ~256 × 256 pixels and smaller. We preprocess the full SEM images with adaptive thresholding[73] to obtain binarized versions of the SEM images with better delineation of the PMMA domains vs. PS domains. We have found adaptive thresholding much better at delineation of PMMA vs. PS domains of SEMs exhibiting the "alternating lines" or the "ladders" morphology compared to other thresholding methods described in **Fig. S2**.

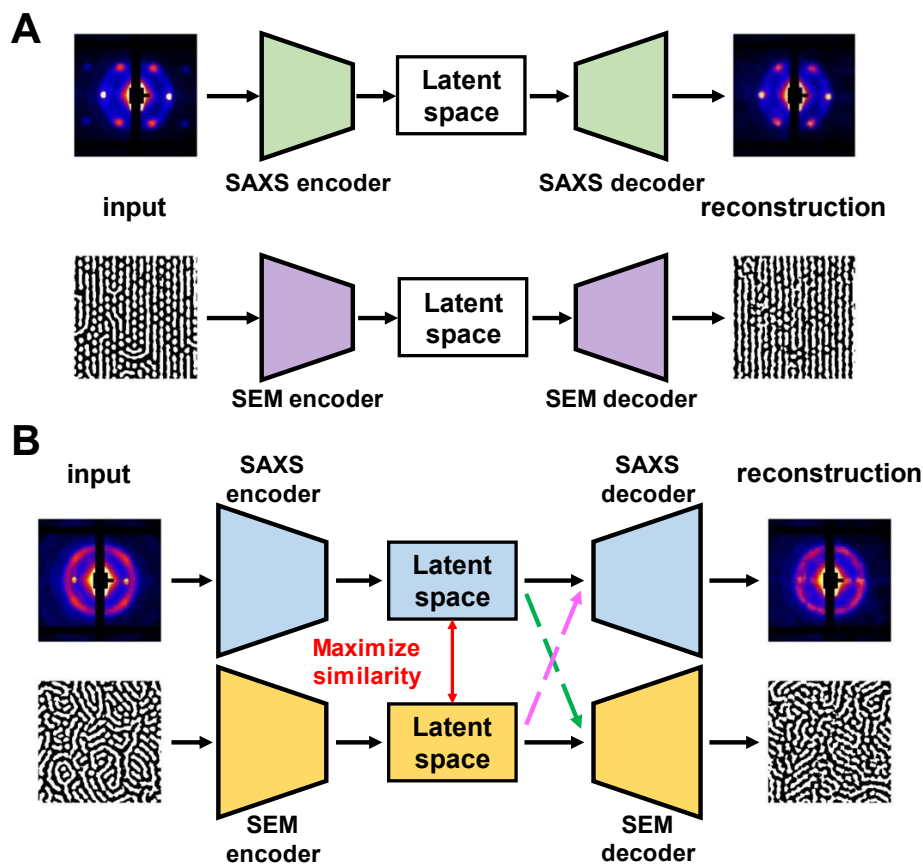

*Fig. 2*. Schematics of our PairVAE workflow. **A.** First, solo training of VAEs for SAXS and SEM images separately. **B.** Pair training of VAEs with SAXS and SEM images concurrently with additional loss functions beyond those used in solo VAEs. PairVAE has four reconstruction capabilities: SAXS profiles from SAXS input (denoted as SAXS→ SAXS in the text), SAXS profiles from SEM input (i.e., SEM→ SAXS), SEM images from SEM input (SEM→ SEM), and SEM images from SAXS input (SAXS→ SEM).

During training of SEM encoder and decoder, we randomly crop a 192 × 192 section of the full SEM at each training epoch and reconstruct the cropped section. Randomly cropping is an effective data augmentation approach for our small-size dataset. For SEM images, we do not expect pixel-to-pixel exact reconstruction from our decoder based on our knowledge of the high statistical variance of image representations belonging to the same morphology. For SEM images, we instead aim for morphologically similar reconstructions of the input image and have applied quantitative metrics to quantify the similarity in morphology: 2-point correlation function and lineal path function. The 2-point correlation function $S(r)$ gives us both the pixel density of the PMMA domains at various positions (radial distance *r*) and the degree of positional order through the oscillations in $S(r)$. The lineal path function tells us the probability distribution of the length of line segments in PMMA domains along a path. If done along the vertical axis, the lineal path function indicates the degree of connectivity of the PMMA domains in the vertical direction; similarly lineal path function can be done along the horizontal path. These two metrics are only used as performance assessment of the reconstructed SEMs (after a model has been trained) and were not calculated or inferred in any way during training. The morphological metrics were calculated with the Porespy package[74]. A lower dimensional latent space is obtained through solo training of each VAE as well as reconstructions of the input characterization data.

In the second part, i.e., pair training, we train the encoders and decoders of the two VAEs in tandem by maximizing the similarity of the two latent spaces as shown in **Fig. 2B**. Additional cross-reconstruction training losses are established to enable cross-reconstruction capabilities of one type of characterization data from another. We note that while solo training is reconstruction of SAXS or SEM in their respective space, pair training attempts to bridge the two different data spaces and tackle the many-to-one (SEM-to-SAXS) and one-to-many (SAXS-to-SEM) cross-reconstruction problems. In selection of training protocols, we find that different from solo training, random cropping of SEM images every epoch during training renders the SEM decoder untrainable as shown by the unabating losses in **Fig. S3.D** and **E**. Therefore, we adopt a two-part training protocol for pair training with a first part of multiple random cropping SEM sessions and a second part focusing on reconstruction on one randomly cropped SEM instance. A detailed description of the VAE models employed, training protocols of both solo VAE training and pair VAE training is in the **Supporting Information**.

## RESULTS and DISCUSSION

First, in **Fig. 3A**, we show the results of solo trained VAE (i.e., a VAE generating SEM image from SEM image and a VAE generating a SAXS pattern from a SAXS pattern). We include one example of the true and reconstructed SAXS pattern and SEM image for each of the four morphologies obtained with solo trained VAE; we note that the selected images are from validation sets and are not seen during training. We observe that the SAXS reconstructions using the solo trained SAXS VAE exhibit a good match with their true counterparts. For SEM reconstructions, we observe only partial resemblance to the true SEM images. For the "skew" and "ladders" morphologies, their SEM reconstructions are visually different than their true SEM images. In **Fig. 3B**, we show the calculated $S(r)$ of PMMA domains along radial distance and the vertical lineal path function of PMMA domains for each true and reconstructed SEM image in **Fig. 3A**. In our case, as the chemical grating direction is vertical, the vertical lineal path function has more features. The horizontal lineal path functions are provided in **Fig. S4**. For all four morphologies' examples, the reconstructed SEMs pixel density of PMMA domains are higher by a small margin with that of the "ladders" morphology being the most different. Reconstructions of both "hexagonal" and "alternating lines" morphologies can capture the spatial order of the PMMA domains indicated as having the similar period of oscillation in $S(r)$ as the $S(r)$ of the true SEM images but give slightly higher lineal path function compared to that of the true SEM images. For the "skew" morphology, the reconstruction does not capture longer range positional order of the true SEM image. As the "ladders" morphology has more statistical variance in the sample, it is not surprising that it is harder to reconstruct "ladders" morphology than the other three morphologies. Additionally, unlike the other three morphologies where the PMMA domains occupy full line segments or dots, in the "ladders" morphology, the PMMA domains reside on the periphery of the ladders. Furthermore, the highly non-uniform shape of the ladders indicates that the domains formed by different ladder peripherals are also non-uniform.

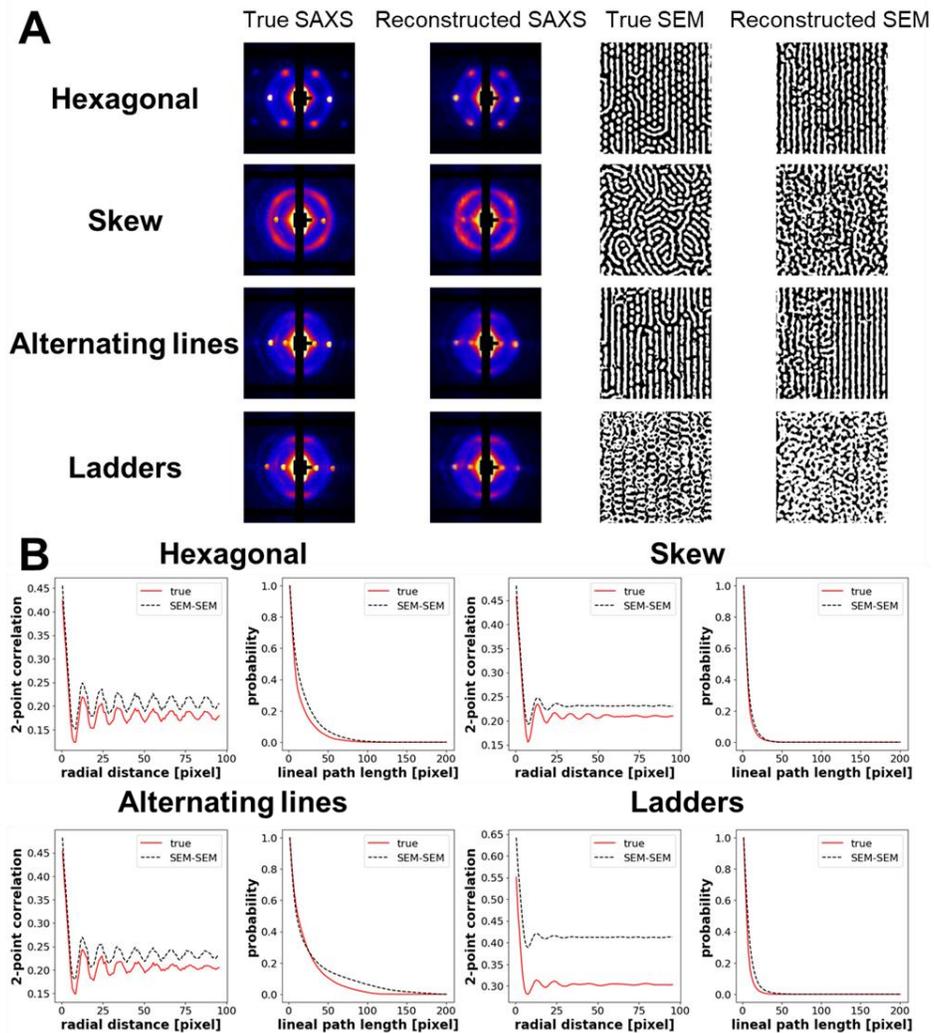

*Fig. 3*. *Reconstruction performance of solo trained VAE. **A.** Examples of SAXS and SEM reconstructions for images from the validation set of the solo trained VAE. **B.** Variation of 2-point correlation functions $S(r)$ of PMMA domains along radial distance and vertical lineal path functions of PMMA domains.*

Next, in **Fig. 4A**, we show the reconstructions from pair trained VAE of the same examples that we showed in **Fig. 3A** (which were obtained with solo trained VAE). We observe in **Fig. 4A** that SAXS reconstruction from SAXS input (denoted as 'SAXS→SAXS') with pair trained VAE is worse than that using solo trained SAXS VAE (**Fig. 3A**). We also observe that SAXS reconstruction from SEM input (denoted as 'SEM→SAXS') is comparable to SAXS reconstruction from SAXS input (i.e., SAXS→SAXS) with pair trained VAE. For SEM reconstruction from SAXS input (denoted as 'SAXS→SEM'), for all four morphologies, the reconstructed SEM images are visually more similar to the true SEM images than the SEM→SEM reconstructions with the pair trained VAE. In general, **Fig. 4A** shows that by pairing SAXS with SEM in pair trained VAE, we have improved the SEM reconstruction and deteriorated SAXS reconstruction as compared to their solo trained VAE (**Fig. 3A**). This unexpected result will be explained below based on latent space complexity in the paragraph after the discussion of results in **Fig. 4B**.

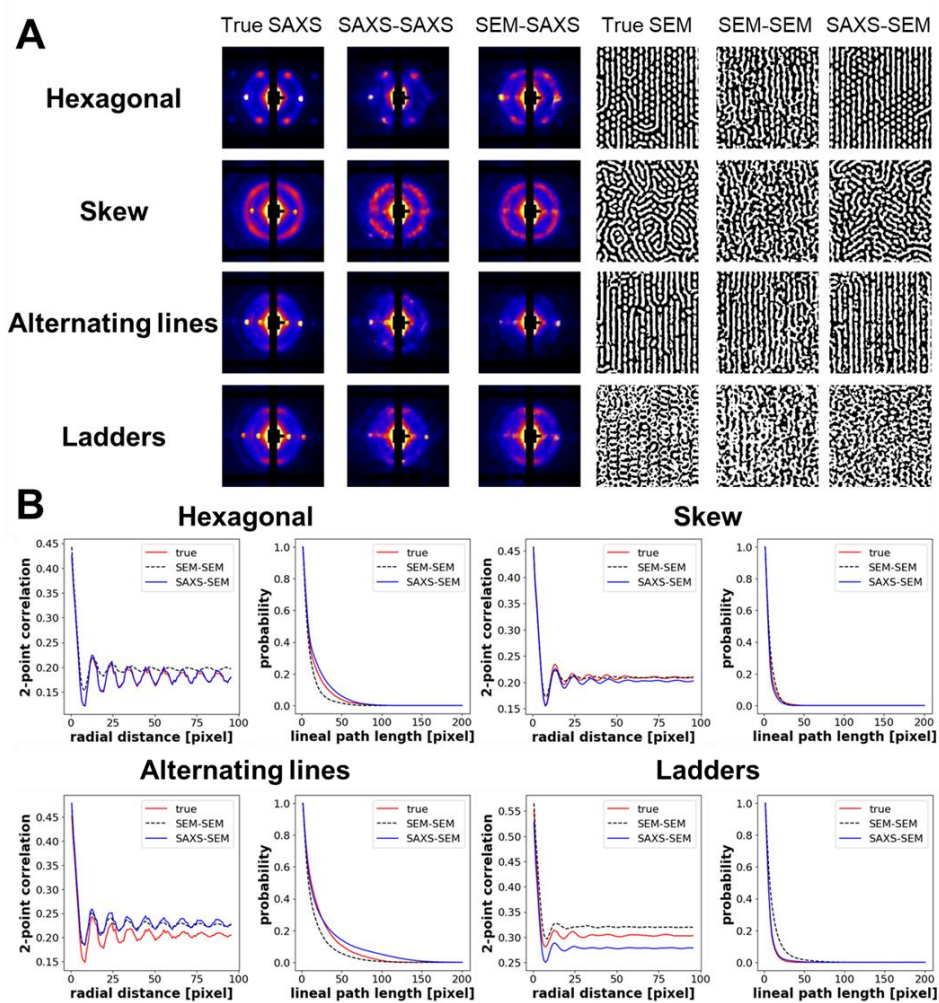

*Fig. 4*. *Reconstruction performance of pair trained VAE.* **A.** *Examples of SAXS and SEM reconstructions for images from the validation set with pair trained VAE.* **B.** *Variation of 2-point correlation functions of PMMA domains along radial distance and vertical lineal path functions of PMMA domains.*

While the conclusions from **Fig. 4A** are purely based on visual comparisons, in **Fig. 4B** we provide a quantitative comparison of the SEM reconstructions against the true SEM for both SEM→SEM and SAXS→SEM, using the two metrics described in the method section – 2-point correlation functions and vertical linear path functions. The horizontal lineal path functions are provided in **Fig. S5**.

For the "hexagonal" morphology, the 2-point correlation function $S(r)$ for SAXS→SEM reconstruction closely matches the $S(r)$ of the true SEM image, while SEM→SEM reconstruction has slightly higher $S(r)$, and does not capture the shape of the $S(r)$ in the true SEM image. Similarly, the vertical lineal path function of SAXS→SEM reconstruction is closer to the true SEM image than the SEM→SEM is, however the vertical lineal path function of SAXS→SEM reconstruction is slightly higher than the true SEM image due to more occurrences of lines than the true SEM image. The SEM→SEM reconstruction's vertical lineal path function is lower than the true SEM image due to more broken-up line segments in the reconstructed image.

For the "skew" morphology, the $S(r)$ of the SAXS→SEM reconstruction follows the oscillations of $S(r)$ of the true SEM image closely albeit with slightly lower values indicating slightly lower pixel density of PMMA domains in the SAXS→SEM reconstruction. The $S(r)$ of the SEM→SEM reconstruction captures the mean of the oscillating $S(r)$ well but does not capture the shape of the $S(r)$. Both SEM→SEM and SAXS→SEM reconstructed "skew" images have lineal path function closely matching that of the true SEM image.

For the "alternating lines" morphology, the $S(r)$ of both SEM→SEM and SAXS→SEM reconstructions have oscillating period like that of the true SEM image. However, both reconstructed $S(r)$ curves are higher than that of the true SEM image. Commenting on the reconstructions of the "alternating lines" morphology in **Fig. 4A.**, the mismatch in the $S(r)$ is likely due to inability to accurately reconstruct the minor dotted region in the true SEM image. The lineal path function of SAXS→SEM reconstruction shows a larger proportion of longer line segments while that of SEM→SEM reconstruction gives smaller probability values overall due to more broken-up line segments.

For the "ladders" morphology, the $S(r)$ of both SEM→SEM and SAXS→SEM reconstructions are still off from that of true SEM image but much better than that of the solo trained SEM reconstruction shown in **Fig. 3B**. The lineal path function of the SAXS→SEM reconstruction is a close match with that of the true SEM image.

The takeaway message so far is that pair training enables cross-reconstruction from either SAXS → SEM or from SEM → SAXS data. Compared with solo trained VAE (results in **Fig. 3**), pair trained VAE (results in **Fig. 4**) improves SEM reconstruction performance while the SAXS reconstruction performance becomes slightly worse. The trade-off in reconstruction performance for the two types of characterization data may be explained by the relative complexity of the two latent spaces. The SAXS data that we use have limited modes or representations making its latent space simpler or more clustered than that of SEM data due to large statistical variance of PMMA domain shapes and positions in SEM images belonging to the same morphology. Through similarity matching in pair training of the SAXS data of simpler or more clustered latent space with SEM data of sparser latent space, the SAXS latent space becomes sparser (leading to worse reconstruction), and the SEM latent space becomes more convergent (leading to better reconstruction).

After the above detailed analysis of selected examples of solo trained VAE and pair trained VAE reconstructions for each morphology, in **Fig. 5**, we show the statistical averages and standard deviations of 2-point correlation function for all SEM true images and reconstructions in **Fig. 5A** for solo trained VAE, and in **Fig. 5B** for pair trained VAE. In **Fig. 5A**, the solo trained VAE's SEM reconstructions give higher 2-point correlation function values on average than the true SEM images. In **Fig. 5B**, the statistical average and variance of 2-point correlation function of both pair trained VAE's SEM reconstructions are a close match with that of the true SEM images. We provide lineal path functions calculated for both solo and pair trained VAEs' reconstructions in **Fig. S6**. We conclude that pair trained VAE improves SEM reconstructions on average for all SEM images as compared to solo trained VAE.

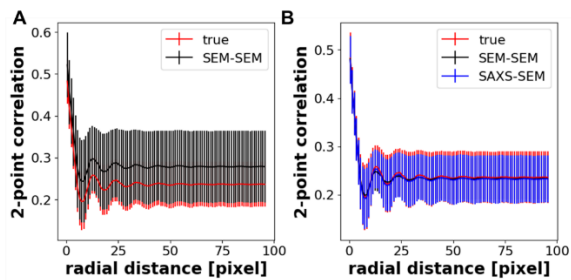

**Fig. 5**. *Statistical average and standard deviation of 2-point correlation functions of PMMA domains of all true SEM (training and validation set) and reconstructions. **A**. From solo trained VAE. **B**. From pair trained VAE.*

We also perform the same solo and pair training on the SAXS and SEM paired dataset but with different number of training set images also with five different selections of the training set and validation set images (determined by a different random seed) at the same training set size to assess the ability to generalize and illustrate the data efficiency of our workflow. We use the mean absolute error (MAE) between the pixel intensity values of a true SAXS image and of a SAXS reconstruction as the performance quantification metric for SAXS reconstruction. We use the MAE between the $S(r)$ of a true SEM image and of a SEM reconstruction as the performance quantification metric for SEM reconstruction. We also use ten random cropped instances from each full SEM image as true SEM images per SAXS image to account for the higher statistical variance of SEM reconstructions as well as for SEM-SAXS reconstructions. The box plots of distributions of all MAEs for SAXS profile reconstructions are shown in **Fig. 6A**. We observe that the median MAE of solo trained SAXS→SAXS reconstruction is much lower than that of the pair trained SAXS→SAXS and SEM→SAXS; this is consistent with what we observed qualitatively from **Fig. 3** to **Fig. 4** that SAXS reconstruction is the best with solo trained VAE. The median MAE of pair trained SEM→SAXS is always a little higher than that of pair trained SAXS→SAXS. We attribute the observed difference to the statistical variability of the SEM images reflecting local morphology that may not be the mode of morphology that the SAXS captures. We also note that the MAEs of all three reconstructions plateau from 48 training set images to 64 training set images implying that we are not improving the reconstruction efficacy that much by going with training sets larger than 48.

The box plots of distributions of all MAEs for SEM image reconstruction are shown in **Fig. 6B**. We observe that the median MAE of pair trained SAXS→SEM reconstruction is always lower than that of the other two reconstructions. The performance difference can be explained by our hypothesis that the more clustered latent space of the SAXS images help converge the SEM latent space to improve SEM reconstruction. The solo and pair trained SEM→SEM reconstructions show comparable MAE for the different training set sizes. SEM reconstructions show improvement with more training set images.

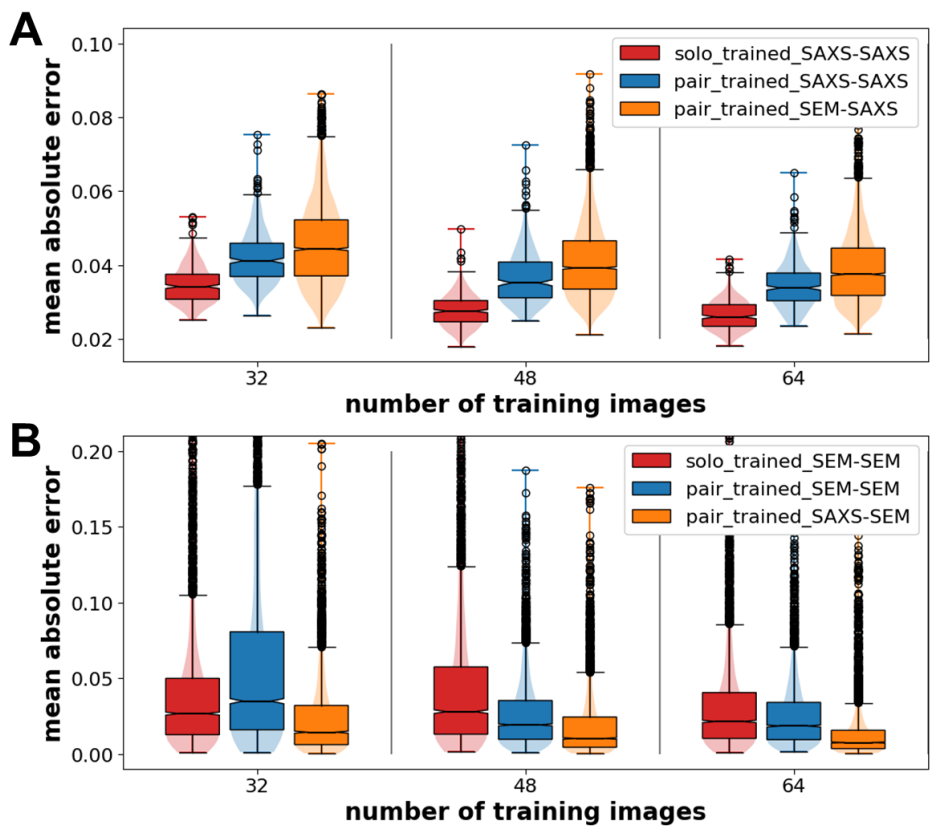

**Fig. 6**. *Solo and pair trained VAE performance assessment using mean absolute error as the metric (vary the number of training images).* ***A.*** *Mean absolute error between each pairing of true and reconstructed SAXS image for all images and multiple VAEs trained with five random seeds, sample size of the boxplot is 360 for solo trained VAE reconstructed SAXS→SAXS and pair trained VAE reconstructed SAXS→SAXS but sample size of the boxplot is 3600 for pair trained VAE reconstructed SEM→SAXS.* ***B.*** *Mean absolute error between each pairing of the 2-point correlation functions of PMMA domains of true and reconstructed SEM images for all images and multiple VAEs trained with five random seeds (we applied ten random crops to each full SEM image during performance assessment), sample sizes of the boxplots are 3600. The median values of both plots are presented in* ***Table S2****. The notch of the boxplots indicates the 95% confidence interval around the median.*

We further probe the capability of PairVAE to generate an ensemble of morphologically similar SEM images directly from the latent space through a trained decoder. We note that we have chosen VAE instead of another deep learning model architecture called U-Net[75] which is the go-to model for biomedical image segmentation. U-Net has proven to be good for segmentation, a pixel-wise reconstruction, owing to the multiple skip connections established between the encoder and decoder at multiple hierarchical levels. And because of the skip connections, the encoder and decoder are trained as a couple in U-Net and cannot be decoupled. Being able to separate the trained encoder and decoder in the VAE model enables us to conduct pair training and explore not just reconstruction but also image generation ability of the decoder.

Here, we demonstrate the image generation ability of the decoder. For this we sample latent variables from multivariate Gaussian distribution constructed with the values of the SAXS latent variables of each morphology example (obtained with pair trained SAXS encoder) shown in **Fig. 4A** as average and variance of the latent variables of the full SAXS dataset as variance. We then use the pair trained SEM decoder to generate images from the sampled latent variables; 5 generated examples are shown in **Fig. 7A**, for each morphology. Generated images present degrees of statistical variations of junctions, vacancies, and defects of PMMA domains that may occur in experimental BCP assembly. When compared to the SAXS→SEM reconstructed images, these generated images (sample size is 100 for each morphology) are morphologically similar as verified visually and by the calculated 2-point correlation function shown in **Fig. 7B**. By changing the magnitude of the variance of the constructed multivariate Gaussian distribution, we observe either better resemblance (smaller variance) **Fig. S7** or less resemblance (larger variance) **Fig. S8** to the SAXS→SEM reconstructions.

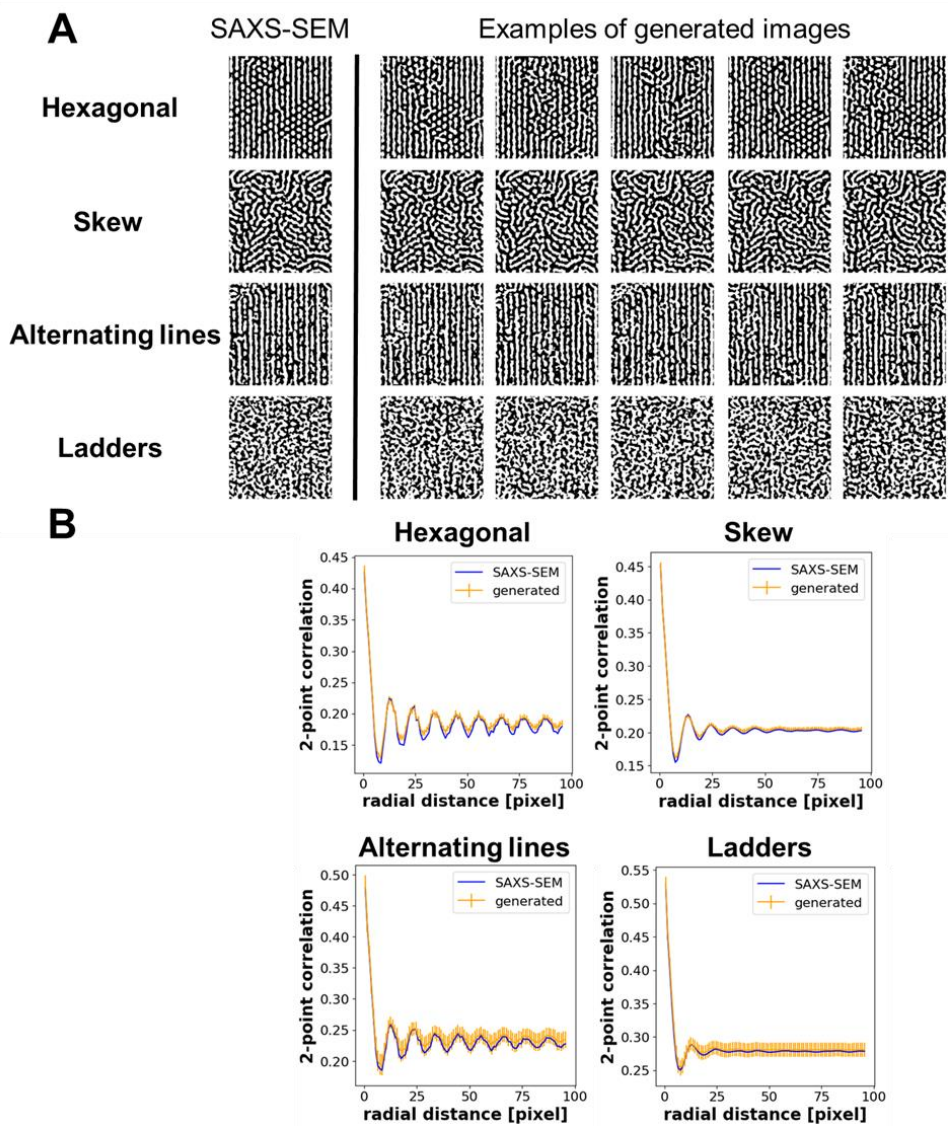

**Fig. 7**. *Performance assessment of generative ability of pair trained VAE's SEM decoder. **A.** SAXS→SEM reconstruction and 5 examples of generated SEM images. **B.** Statistical average and standard deviations of 2-point correlation functions of PMMA domains of 100 generated SEM images for each morphology example plotted against that of the SAXS→SEM reconstruction.*

## CONCLUSION

In conclusion, we have developed a novel machine learning flow called 'PairVAE' that enables cross-reconstruction capabilities for complementary material structural characterization data. Here we demonstrated a proof-of-concept application of PairVAE using complementary characterization data - 2D SAXS patterns and SEM images - of ordered block copolymer morphology. By pairing up the latent spaces of SAXS and SEM, PairVAE enables cross-reconstruction of SAXS profiles from SEM input or SEM images from SAXS input. We were able to generate morphologically similar SEM images from SAXS input with close match of the 2-point correlation functions of the PMMA domains to that of the true SEM images for all data on average. We provide an explanation for the performance trade-off of pair trained VAE seen for SAXS reconstructions (worse than solo trained VAE) and SEM reconstructions (better than solo trained VAE). The explanation is that by pairing up SEM latent space (relatively sparse) with SAXS latent space (relatively clustered), the SEM latent space becomes more convergent, yielding morphologically closer reconstructions than solo training, while SAXS latent space becomes less clustered, yielding more unphysical SAXS patterns than solo training. Our PairVAE implementation for SAXS-SEM reconstruction specifically incorporated random cropping of SEM images during training as a means of data augmentation that help mitigate the small-data handicap. Lastly, we have also shown that generation of ensembles of morphologically similar SEM images is possible by sampling latent variables directly; this is possible only because we can decouple the trained encoder and decoder which is not possible with more commonly used image segmentation models like U-Net[75].

The impact of this PairVAE workflow is in the many potential applications it offers, some of which we list below:

1) Generation of bottleneck characterization data (e.g., laborious sample preparation for microscopy) from high throughput data (e.g., scattering profiles collection) that have complementary purposes.
2) Generation of easier-to-interpret characterization data (e.g., microscopy image) from harder-to-interpret characterization data (e.g., scattering data) in a data-driven manner without intermediate manual interpretation.
3) Generation of ensembles of structure or morphology data with designed morphological features for downstream calculation or simulation of other physical properties such as optical property, electrical conductivity property or rheological properties.
4) Integration with time series data for prediction of time evolution of structure or morphology or other characterization data or properties.
5) (For scattering and microscopy specifically) Direct integration with microscopy and scattering experimental setups for real-time high throughput multi-purpose characterization of material structure or morphology, or in-situ collection and generation of movies of structural characterizations.

## ASSOCIATED CONTENT

**Supporting information**

Description of the block copolymer SAXS-SEM morphology characterization dataset, image data preprocessing procedures, python packages utilized and the usages of each package, the variational autoencoder model, the solo training protocol, the pair training protocol, and additional lineal path function figures. Supporting information is available from the corresponding author upon request.

## AUTHOR INFORMATION

**Corresponding Author**

**Arthi Jayaraman** – Department of Chemical and Biomolecular Engineering and Department of Materials Science and Engineering, University of Delaware, Newark, Delaware 19716, United States

**Author**

**Shizhao Lu** - Department of Chemical and Biomolecular, University of Delaware, Newark, Delaware, 19716, United States

**Acknowledgements:** The authors acknowledge financial support from the U.S. Department of Energy, Office of Science Grant Number DE-SC 0023264. The authors thank Gregory Doerk, Kevin Yager, and their co-authors for making their block copolymer characterization data openly available. The authors thank Gregory S. Doerk, Quentin Michaudel, Ryan C. Hayward, Jonathan A. Malen, Austin M. Evans, Brent S. Sumerlin, Pramod Reddy, Vasudevan Venkateshwaran, and Rama K. Vasudevan for their feedback and discussion regarding our workflow and results.

**Notes**

The authors declare no competing financial interest.